\documentclass[longbibliography,groupedaddress,showpacs,showkeys,amssymb,eqsecnum,aps,nofootinbib]{revtex4}
\usepackage[pdftex]{graphicx}
\usepackage[pdftex]{graphics}
\usepackage{amsmath} 
\usepackage{epsf}                                                                                           
\usepackage{color}                   
\usepackage{verbatim}                                                                         
\usepackage{hyperref}

\newcommand{\be}{\begin{equation}}

\newcommand{\ee}{\end{equation}}

\begin{document}                                                                                              

\title{Consistency Conditions for the First-Order Formulation of Yang-Mills Theory}

\author{D. G. C. McKeon}
\email{dgmckeo2@uwo.ca}
\affiliation{
Department of Applied Mathematics, The University of Western Ontario, London, Ontario N6A 5B7, Canada}
\affiliation{Department of Mathematics and Computer Science, Algoma University,
Sault Ste.~Marie, Ontario P6A 2G4, Canada}

\author{F. T. Brandt}  
\email{fbrandt@usp.br}
\affiliation{Instituto de F\'{\i}sica, Universidade de S\~ao Paulo, S\~ao Paulo, SP 05508-090, Brazil}

\author{J. Frenkel}
\email{jfrenkel@if.usp.br}
\affiliation{Instituto de F\'{\i}sica, Universidade de S\~ao Paulo, S\~ao Paulo, SP 05508-090, Brazil}

\author{S. Martins-Filho}    
\email{sergiomartinsfilho@usp.br}
\affiliation{Instituto de F\'{\i}sica, Universidade de S\~ao Paulo, S\~ao Paulo, SP 05508-090, Brazil}

\date{\today}

\begin{abstract}
We examine the self-consistency of the first-order formulation of the
Yang-Mills theory. By comparing the generating functional $Z$ before
and after integrating out the additional field $F^a_{\mu\nu}$, we
derive a set of structural identities that must be satisfied by the
Green's functions at all orders. These identities, which hold in any
dimension, are distinct from the usual Ward identities and are
necessary for the internal consistency of the first-order
formalism. They relate the Green's functions involving the fields
$F^a_{\mu\nu}$, to Green's functions in the second-order formulation
which contain the gluon strength tensor $f^a_{\mu\nu}$.
In particular, such identities
may provide a simple physical interpretation 
of the additional field  $F^a_{\mu\nu}$.
\end{abstract}                                                                                                

\pacs{11.15.-q}
\keywords{1st and 2nd order gauge theories; renormalization; composite fields}

\maketitle

\section{Introduction} 
The first-order formulation of gauge theories has a simple form
that involves
only cubic interactions of the gauge fields, which are
momentum-independent. This simplifies the computations of the quantum
corrections in the standard second-order gauge theories, which involve
momentum-dependent three-point as well as higher-point vertices. 
It is well known that the first-order formulation may be achieved 
by introducing, for example in the Yang-mills theory, 
an auxiliary field $F^a_{\mu\nu}$  
\cite{Okubo:1979gt,McKeon:1994ds,Martellini:1997mu,costello:2011b,Brandt:2015nxa,Brandt:2016eaj,Frenkel:2017xvm,Frenkel:2018xup,Brandt:2018wxe}.
The corresponding first-order 
Lagrangian density may be written as 
\be
\tilde{\cal L} =  \frac 1 4 F_{\mu\nu}^a F^{\mu\nu\,a}-\frac 1 2 f_{\mu\nu}^a F^{\mu\nu\,a}
\ee
where $f_{\mu\nu}^a$ is the gluon field strength tensor
\be
f_{\mu\nu}^a = \partial_\mu A^a_\nu - \partial_\nu A^a_\mu +g f^{abc} A^b_\mu A^c_\nu.
\ee
Using the Euler-Lagrange equation in conjunction with the above Lagrangian, one can see that,
at the classical level, $F^a_{\mu\nu}=f^a_{\mu\nu}$.
From this it follows 
that $\tilde{\cal L} = -1/4 f_{\mu\nu}^a f^{a\,\mu\nu}$,
which corresponds to the usual second-order Lagrangian.

At the quantum level, the renormalization of
the first-order formalism has been previously studied from various points
of view
\cite{Okubo:1979gt,McKeon:1994ds,Martellini:1997mu,costello:2011b, 
Brandt:2015nxa,Brandt:2016eaj,Frenkel:2017xvm,Frenkel:2018xup,Brandt:2018wxe}.
In particular, the BRST renormalization
of this formulation has been addressed in \cite{Frenkel:2017xvm,Frenkel:2018xup}.
The BRST identities, which reflect
the gauge invariance of the theory, are suitable for a recursive proof of the renormalizability to all orders in perturbation theory
\cite{Buchbinder:2018jqs,Lavrov:2020exa,Barvinsky:2017zlx}.

In the present work, we examine different kinds of identities, which are necessary for the consistency of the first-order formulation. To this end, we introduce a source $j^a_\mu$ for the gluon
field $A^a_\mu$ and also a source $J^a_{\mu\nu}$  for the field $F^a_{\mu\nu}$, and consider
the generating functional $Z[J,j]$ of  Green's functions. We compare the functional dependence of $Z$ on the sources in the original first-order formalism with that found after making a shift
\be\label{eq3}
F_{\mu\nu}^a\rightarrow F_{\mu\nu}^a + f_{\mu\nu}^a - 2 J_{\mu\nu}^a
\ee
which enables us to integrate out the auxiliary field
$F_{\mu\nu}^a$. The equality of these functional forms leads to a set
of structural identities among the Green's functions which must be
satisfied to all orders, in any dimension. These show that in the
first-order formalism, 
Green's functions containing only external gluon fields are the same
as those which occur in the second-order formulation. Furthermore,
these identities relate the Green's functions with some external  
$F_{\mu\nu}^a$ 
fields to certain Green's functions in the second-order formalism 
that contain the gluon strength 
tensor field $f_{\mu\nu}^a$. 
Such Green's functions involve composite fields,
in which the external
legs are ``pinched'' at the same spacetime point. As is well known, 
these lead to ultraviolet (short-distance)
singularities \cite{Wilson:1972ee,muta:book87,weinberg:book1995}.
In our case, such singularities play an important role. They
are essential for the cancellation of ultraviolet divergences arising from loop diagrams,
which is necessary for the implementation of the structural
identities. Proceeding in this way, one obtains, in particular, the identity
\be\label{eq4}
\langle 0|T F_{\mu\nu}^a(x)A^{b_1}_{\alpha_1}(x_1)\cdots
A^{b_n}_{\alpha_n}(x_n) |0\rangle  =
\langle 0|T f_{\mu\nu}^a(x)A^{b_1}_{\alpha_1}(x_1)\cdots A^{b_n}_{\alpha_n}(x_n) |0\rangle. 
\ee
This may be interpreted as the quantum-mechanical generalization 
of the relation $F_{\mu\nu}^a = f_{\mu\nu}^a$ obtained at the classical level. Namely, the Green's functions containing one $F_{\mu\nu}^a$ field and an arbitrary number of Yang-Mills fields are equal to the Green's functions involving a single gluon field strength tensor $f_{\mu\nu}^a$ and an arbitrary number of Yang-Mills fields.
    
In section 2 we analyse the Lagrangian and the generating functional of Green's
functions in the first-order formulation. 
In section 3, we proceed to derive the result 
\eqref{eq4}, which has been explicitly verified up to one-loop 
order. 
In section 4, we study a
basic structural identity satisfied by the Green's functions and
examine the cancellations between the loop ultraviolet 
divergences and the ultraviolet singularities arising from tree 
graphs involving composite fields. 
We present a brief discussion of the results
in section 5.
Some details of the relevant calculations are given in the 
Appendices.

\section{The Lagrangian and the generating functional}
The complete Lagrangian density for the first-order formulation in
covariant gauges is
\be\label{eq5}
{\cal L} =  \frac 1 4 F_{\mu\nu}^a F^{\mu\nu\,a}-\frac 1 2 f_{\mu\nu}^a F^{\mu\nu\,a}
-\frac{1}{2\xi} \left(\partial_\mu A^{\mu\, a}\right)^2+
\left(\partial_\mu {\bar\eta}^a\right) D^{\mu\, ab}\eta^b,
\ee
where $\xi$ is a gauge fixing parameter, ${\bar\eta}^a$, $\eta^b$ are
ghost fields and $D^{\mu\, ab}$ is the covariant derivative
\be\label{eq6}
D^{\mu\, ab} = \delta^{ab} \partial^\mu - g f^{abc} A^{\mu\, c}.
\ee
In addition, we will also introduce the external sources
$J^a_{\mu\nu}$ and $j^a_{\mu}$ as follows
\be\label{eq7}
{\cal L}_{source} = J^a_{\mu\nu} F^{\mu\nu\, a} + j^a_\mu A^{\mu\, a}.
\ee
The normalized generating functional for Green's functions is given by
\be\label{eq8}
Z[J,j] = \frac{\int {\cal D}\eta {\cal D}\bar\eta {\cal D} F {\cal D}A 
\exp i \left[S+\int d^d x\left(J^a_{\mu\nu} F^{\mu\nu\,a} + j^a_{\mu} A^{\mu\,a}\right)\right]}
{\int {\cal D}\eta {\cal D}\bar\eta {\cal D} F {\cal D}A \exp i S},
\ee
where $S=\int d^d x {\cal L}$. This equation is in a form suitable for
functional differentiation with respect to $J$ and $j$, and therefore
for finding the Green's functions.

If we were to set $J_{\mu\nu}^a=0$ at the outset
(so that we would consider Green's functions with only external fields
$A_\mu^a$) and make the change of variable in the functional integral
\be\label{eq9}
F_{\mu\nu}^a\rightarrow F_{\mu\nu}^a + f_{\mu\nu}^a
\ee
then we find that
\be\label{eq10}
Z[J=0,j] = Z_2[j],
\ee
where $Z_2[j]$ is the generating functional for the second-order
theory, characterized by the Lagrangian density
\be\label{eq11}
{\cal L}_2 = -\frac 1 4 f_{\mu\nu}^a f^{\mu\nu\,a}
-\frac{1}{2\xi} \left(\partial_\mu A^{\mu\, a}\right)^2+
\left(\partial_\mu {\bar\eta}^a\right) D^{\mu\, ab}\eta^b
\ee
together with the source term $j^a_\mu A^{\mu\, a}$.
This establishes the important property that the Green's functions
with only external gluon fields are the same in both approaches.

We now consider using  $Z[J,j]$ with $J\neq 0$ and examine what
changes occur in the first-order formalism when there are external
fields $F_{\mu\nu}^a$. To this end we make the shift \eqref{eq3} 
in the numerator of \eqref{eq8}
which leads, after integrating out
the $F$ field, to the alternative form of the generating functional
\be\label{eq12}
Z^\prime[J,j] = \frac{\int {\cal D}\eta {\cal D}\bar\eta  {\cal D}A 
\exp i \int d^d x\left({\cal L}_2+J^a_{\mu\nu} f^{\mu\nu\,a}-J^a_{\mu\nu} J^{\mu\nu\,a} 
+ j^a_{\mu} A^{\mu\,a}\right)}
{\int {\cal D}\eta {\cal D}\bar\eta  {\cal D}A \exp i \int d^d x{\cal L}_2}.
\ee
This equals to $Z_2[j]$ in Eq. \eqref{eq10} if we set 
$J_{\mu\nu}^a = 0$.  It is worth noticing  here the unusual dependence
of  $Z^\prime[J,j]$ on $J_{\mu\nu}^a$. 

Comparing the forms \eqref{eq8}
and \eqref{eq12} of the generating functionals and differentiating
these with respect to $J$ and $j$, leads to a set of structural identities among the
Green's functions, which must be satisfied to all orders. In
principle, the Green's functions obtained in this way should be
evaluated by using the Feynman rules appropriate to the
first-order formalism. However, since Green's functions with only
external gluon fields are the same as those in 
the second-order formulation, the Green's functions obtained via  
Eq. \eqref{eq12} are equal to the corresponding 
ones obtained by using this formulation. Therefore, we see 
that such structural identities may relate the Green's functions 
involving  some $F$ fields to certain Green's  functions 
in the second-order formulation that contain gluon strength tensor 
fields $f_{\mu\nu}^a$. 
These identities hold in any dimensions, both for the finite as well
as for the ultraviolet divergent parts of the Green's functions.

\section{Derivation of relation (\ref{eq4}) }


Taking the functional differentiation of
Eqs. \eqref{eq8} and \eqref{eq12}  
with respect to $J$ and $j$, and equating the results we obtain,
by setting $J = j = 0$, the equation
\be\label{eq19}
\langle 0| T F_{\mu\nu}^a(x) A_\alpha^b(y) | 0 \rangle =
\langle 0| T f_{\mu\nu}^a(x) A_\alpha^b(y) | 0 \rangle . 
\ee
Using the Feynman rules given in Appendix A one can verify 
that this equation, which relates the propagators $F A$ and $f A$, 
is satisfied in the tree approximation. To one-loop
order, the divergent part of the left hand side  in momentum space is  
(see Eq. \eqref{propFAUV})
\be\label{propFAUVt}
{D^{(1)}_{FA}}_{\mu \nu , \alpha}^{ab} =
\frac{C_{YM} g^2}{16\pi^2\epsilon}
\delta^{ab}
\frac{11-9\xi}{12}\frac{1}{k^2}
\left(k_\mu \eta_{\nu\alpha}-k_\nu \eta_{\mu\alpha}\right) ,
\ee
where we have used dimensional regularization in $d=4-2\epsilon$
dimensions.
Our conventions are such that the configuration space 
derivative $\partial_\mu$ becomes in momentum space 
$+i k$ where the momentum $k$ is flowing into the vertex
with which it is associated.

\begin{figure}[t]
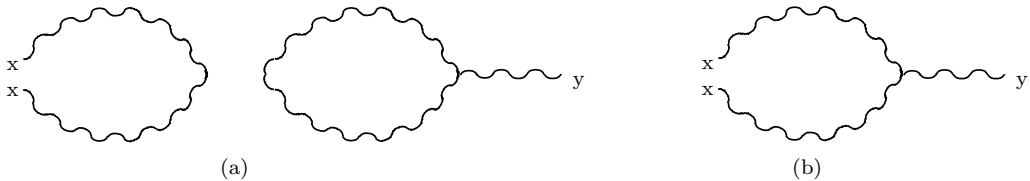

\begin{eqnarray}
\input{pinch_tad1.pspdftex} & \qquad \qquad  
\input{pinch_tad2.pspdftex} 
 \nonumber 
\end{eqnarray}
\caption{Tadpole (a) and ``pinched'' contributions (b) from
    $\langle 0|T  f^{acd} A_\mu^c(x) A_\nu^d(x)  A_\alpha^b(y)  |0\rangle$.} 
\label{fig1}
\end{figure}

We must now calculate the divergent part of  the propagator on the
right hand side of Eq. \eqref{eq19}. One contribution to this comes 
from the 
$\partial_\mu A_\nu^a -\partial_\nu A_\mu^a$ part of $f_{\mu\nu}^a$.
With the help of the Eq. \eqref{dAAUV}, this part yields 
in momentum space
\be\label{eq21}
\frac{C_{YM} g^2}{16\pi^2\epsilon}\delta^{ab} 
\frac{13-3\xi}{6}
\frac{1}{k^2}
\left(k_\mu \eta_{\nu\alpha} - k_\nu \eta_{\mu\alpha} \right) .
\ee
The other contribution comes from the composite field
$g f^{abc} A_\mu^b(x) A_\nu^c(x)$
which occurs in $f_{\mu\nu}^a(x)$.
Using Wick's theorem, one can verify that to order $g^2$, such a term
arises from the Feynman diagrams shown in Fig. \ref{fig1}.             
The first graph contains a tadpole, Fig (\ref{fig1}a),
which vanishes by using dimensional regularization.
The second diagram corresponds to a three-point tree Green's function
which has however two coordinates ``pinched'' at the same spacetime
point $x$. As noted earlier, such a composite field leads to an
ultraviolet (short-distance) singularity (see also Appendix B).
Using the well known
expression for the three-point gluon vertex, it is straightforward to
evaluate this contribution in momentum space, which turns out to be
(see Eq. \eqref{eqAAAfAUV})
\be\label{eq22}
-\frac{C_{YM} g^2}{16\pi^2\epsilon} \delta^{ab}  \frac{\xi+5}{4} 
\frac{1}{k^2}
\left(k_\mu \eta_{\nu\alpha}-k_\nu \eta_{\mu\alpha}\right) .
\ee 
Adding the contributions given in equations \eqref{eq21} and
\eqref{eq22}, we obtain a result which agrees with that 
given in Eq. \eqref{propFAUVt}.
Thus, we have explicitly verified, to one-loop order,
the validity of the identity \eqref{eq19} for the UV divergent parts
(in Appendix B we show that this is valid for the full expression in
$d$ dimensions).
It is straightforward to generalize Eq. \eqref{eq19}
so as to include an arbitrary number of gluon fields, namely
\be\label{eq23}
\langle 0|T F_{\mu\nu}^a(x)A^{b_1}_{\alpha_1}(x_1)\cdots A^{b_n}_{\alpha_n}(x_n) |0\rangle=
\langle 0|T f_{\mu\nu}^a(x)A^{b_1}_{\alpha_1}(x_1)\cdots A^{b_n}_{\alpha_n}(x_n) |0\rangle. 
\ee
As we have mentioned, this relation may be interpreted as being a quantum-mechanical
extension of the relation $F_{\mu\nu}^a=f_{\mu\nu}^a$ which holds at the classical level.

\section{A basic structural identity}
Applying $\delta^2/\delta J_{\mu\nu}^a(x)\delta J_{\alpha\beta}^b(y)$
to Eqs. \eqref{eq8} and \eqref{eq12} and equating the results, leads
to
\be\label{eq13}
\langle 0| T F_{\mu\nu}^a(x) F_{\alpha\beta}^b(y) | 0 \rangle =
2 i\delta^{ab} I_{\mu\nu ,\alpha\beta}\delta^4(x-y) +
\langle 0| T f_{\mu\nu}^a(x) f_{\alpha\beta}^b(y) | 0 \rangle , 
\ee
where $I_{\mu\nu , \alpha\beta}$ is given by Eq. \eqref{eqAI}.
As we have explained following Eq. \eqref{eq12}, the above equation
relates the propagators $\langle 0| T F_{\mu\nu}^a(x) F_{\alpha\beta}^b(y) | 0 \rangle$
calculated in the first-order formalism to the Green's functions
$\langle 0| T f_{\mu\nu}^a(x) f_{\alpha\beta}^b(y) | 0 \rangle$
computed in the second-order formalism. We now will examine the
perturbative expansion of each side of Eq. \eqref{eq13}. It is easy to
verify that this equation is satisfied at the tree level. To one-loop
order, using the Feynman rules given in Appendix A, one can show
(see Eq. \eqref{propFFUV}) that the graphs that contribute to the left
side (see Fig. \ref{fig4}) yield in momentum space the pole term
\be\label{eq14}
{D^{(1)}_{FF}}_{\mu \nu , \alpha\beta}^{ab}  =
i \frac{C_{YM} g^2}{16\pi^2\epsilon} \delta^{ab}
\left\{
-(\xi+1) I_{\mu\nu, \alpha\beta}+
\left(\frac{2}{3} + 2 \xi\right) 
\frac{1}{k^2}L_{\mu\nu,\alpha\beta}(k) 
\right\} + \cdots ,
\ee
where $L_{\mu\nu ,\alpha\beta}(k)$ is given by Eq. \eqref{eqAL}
and we use dimensional regularization in $4-2\epsilon$ dimensions.

The computation of the right hand side of Eq. \eqref{eq13} is somewhat
more involved at order $g^2$, where we encounter contributions from three
sources. The first one, which corresponds to 
$\langle 0| T \left(\partial_\mu A_\nu^a -\partial_\nu A_\mu^a\right)(x) 
\left(\partial_\alpha A_\beta^b -\partial_\beta A_\alpha^b\right)(y) | 0 \rangle$
comes from one-loop graphs shown in Fig. \ref{fig2}. This yields in momentum
space the pole term (see Eq. \eqref{dAdAUV}) 
\be\label{eq15}
i \frac{C_{YM} g^2}{16\pi^2\epsilon}\delta^{ab}  
\frac{3\xi-13}{3} 
\frac{1}{k^2}  L_{\mu\nu,\alpha\beta}(k). 
\ee
The other contributions arise from the composite fields which
occur in the Green's functions 
$g f^{bcd} \langle 0|T(\partial_\mu A_\nu^a-\partial_\nu A_\mu^a)A_\alpha^cA_\beta^d |0\rangle$, $g f^{acd} \langle 0|TA_\mu^cA_\nu^d(\partial_\alpha A_\beta^b-\partial_\beta A_\alpha^b) |0 \rangle$ and $
g^2 f^{acd}f^{bc'd'} \langle 0|TA_\mu^cA_\nu^dA_\alpha^{c'}A_\beta^{d'}|0 \rangle$.
%
Since the tadpole graphs vanish when using dimensional regularization,
the only Feynman diagrams which contribute to these Green's functions
are shown in Figs. (\ref{fig5}a),  (\ref{fig5}b) and (\ref{fig5}c)
respectively. 
These yield, in momentum
space, the following pole terms (see Eqs. \eqref{fAAUV} and 
\eqref{AAAAUV}  ) 
\be\label{eq16}
i \frac{g^2 C_{YM}}{16 \pi^2 \epsilon} \delta^{ab}\left(\xi+5\right) 
\frac{1}{k^2} L_{\mu\nu ,\alpha\beta}(k)  
\ee
and
\be\label{eq17}
-i\frac{g^2 C_{YM}}{16 \pi^2 \epsilon} \delta^{ab} (\xi+1)
I_{\mu\nu ,\alpha\beta}. 
\ee

There is an aspect of the contributions from Fig. \ref{fig5} that is
worth pointing out. The divergent terms given respectively by
Eqs. \eqref{eq16} and \eqref{eq17}, come from ``pinching'' at the same
spacetime point the legs of what would otherwise be a tree
diagram. Explicit calculation of such diagrams once the external legs
are ``pinched'', gives rise to short distance (ultraviolet) singularities
as $\epsilon\rightarrow 0$ when using dimensional
regularization. Adding the contributions coming from
Eqs. \eqref{eq15},  \eqref{eq16} and \eqref{eq17} leads to the result
\eqref{eq14}, thereby verifying the pole part of the 
identity \eqref{eq13} to order $g^2$
(in Appendix B we show that this is valid for the full expression in
$d$ dimensions).

%

\section{Discussion}
We have studied certain consistency conditions for the first-order
formulation of the Yang-Mills theory. To this end, we examined 
the forms of the generating functionals  of Green's 
functions $Z (J,j)$, before and after integrating out the additional
field $F_{\mu\nu}^a$. Differentiations of these forms with respect to 
$J_{\mu\nu}^a$ and $j_{\mu}^a$ yield a set of structural identities 
which are complementary but distinct from the usual Ward identities.
Such identities lead to connections between the Green's functions 
involving the field $F_{\mu\nu}^a$     
and the Green's functions in the second-order formulation 
that contain the gluon strength tensor $f_{\mu\nu}^a$.      
An interesting outcome of these relations is a quantum-mechanical 
extension of the classical result $F_{\mu\nu}^a =f_{\mu\nu}^a$, which 
provides a simple interpretation of the field $F_{\mu\nu}^a$.
       
The structural identities hold for the complete Green's  functions, in
any dimensions and to all orders. We have explicitly verified such
identities to one loop-order, for the ultraviolet divergent parts.
These require subtle cancellations between the ultraviolet divergences 
coming from loop graphs and the short-distance singularities 
induced by the composite fields present in the gluon strength tensor
$f_{\mu\nu}^a$. These results provide a simpler 
computation of the expectation 
values of  time-ordered products of operators containing 
the composite gluon strength tensor
$f^a_{\mu\nu}$, in terms of those involving 
the local field $F^a_{\mu\nu}$.

It is known that the renormalizability of the 
first-order formulation requires, as well as a scaling of the
$F_{\mu\nu}^a$ field, also a mixing with the 
gluon strength tensor field $f_{\mu\nu}^a$ 
\be\label{eq24}
F_{\mu\nu}^a\rightarrow Z_F^{1/2} F_{\mu\nu}^a +Z_{F f} f_{\mu\nu}^a 
\ee
where $Z_{F f}$ is a counter-term which is equal to  
$(1- 3\xi) g^2 C_{YM}/192\pi^2\epsilon$, at one-loop order \cite{Frenkel:2017xvm}.
Hence, one may also expect a scaling and mixing of 
sources of the form
\be\label{eq25}
j_{\mu}^a\rightarrow Z_j^{1/2} j_{\mu}^a + z D^{\nu\,ab} J_{\mu\nu}^b 
\ee
which is admissible on dimensional, Lorentz and charge-conjugation
symmetry grounds. Yet, our explicit one-loop calculations show that
$z=0$. This result may be understood by noting that the last term in
Eq. \eqref{eq25} could induce corrections which would violate the
Eq. \eqref{eq13}. Thus we infer that, to all orders, the
structural identities forbid a mixing between the 
sources $j_\mu^a$ and $J_{\mu\nu}^a$. 

Finally, we remark that the first-order formalism is also useful 
in quantum gravity, where it allows us to replace an 
infinite number of complicated multiple graviton couplings by a 
finite number of simple cubic vertices 
\cite{Brandt:2015nxa,Brandt:2016eaj}. 
In this theory, one would similarly get
corresponding structural identities, which ensure the internal
consistency of such a formulation.This is an interesting issue which deserves further study.

\begin{acknowledgments}
F. T. B. and J. F. thank CNPq (Brazil) for financial 
support. S. M-F thanks CAPES (Brazil) for financial support. 
D. G. C. M. thanks Roger Macleod for an enlightening discussion. 
This work comes as an aftermath of an original project developed with 
the support of FAPESP (Brazil), grant number 2018/01073-5. 
\end{acknowledgments}

%

\newpage
 
\appendix

\section{Feynman rules}
The following Feynman rules for the first-order Yang-Mills theory 
can be readily obtained from the Lagrangian density in Eq. \eqref{eq5}
(for details, see Ref. \cite{Brandt:2015nxa})
\begin{subequations}\label{eqA1}
\begin{eqnarray}
\begin{array}{c}\label{eqA1a} 
\input{Fprop.pspdftex} 
\end{array} \;\;\;\;  &\;\;& 
2 i\left(I_{\lambda\sigma,\rho\kappa}-\frac{1}{p^2} L_{\lambda\sigma,\rho\kappa}(p)\right)\delta^{ab},
\\
\nonumber 
\\
\begin{array}{c}\label{eqA1b}
\input{Aprop.pspdftex} 
\end{array} \;\;\;\;  &\;\;& 
-\frac{i}{p^2}\left( \eta_{\mu\nu} -\frac{1-\xi}{p^2}p_\mu p_\nu\right)\delta^{ab}, 
\\
\nonumber 
\\
\begin{array}{c}\label{eqA1c} 
\input{AF.pspdftex} 
\end{array} \;\;\;\;  &\;\;& 
-\frac{1}{p^2}\left(p_\rho \eta_{\kappa\mu} - p_\kappa\eta_{\rho\mu}\right)\delta^{ab},
\\
\nonumber 
\\
\begin{array}{c}\label{eqA1d} 
\input{FA.pspdftex} 
\end{array} \;\;\;\;  &\;\;& 
\frac{1}{p^2}\left(p_\lambda \eta_{\sigma\nu} - p_\sigma\eta_{\lambda\nu}\right)\delta^{ab},
\\
\nonumber 
\\
\begin{array}{c}\label{eqA1e} 
\input{FAA.pspdftex} 
\end{array} \;\;\;\;  &\;\;& 
-i\frac{g}{2}f^{abc}\left(\eta_{\lambda\mu} \eta_{\sigma\nu} - \eta_{\sigma\mu}\eta_{\lambda\nu}\right), 
\\
\nonumber 
\\
\begin{array}{c}\label{eqA1f} 
\input{ghost.pspdftex} 
\end{array} \;\;\;\;  &\;\;& 
\frac{i}{p^2}\delta^{ab}, 
\\
\nonumber 
\\
\begin{array}{c}\label{eqA1g} 
\input{Acc.pspdftex} 
\end{array} \;\;\;\;  &\;\;& 
g f^{abc} p_\mu , 
\end{eqnarray}
\end{subequations}
where the quanta of the $A_\mu^a$ and $F_{\mu\nu}^a$ fields are represented respectively  by the wavy and the continuous lines. Here, 
the tensors $I_{\mu\nu , \alpha\beta}$ and $L_{\mu\nu ,\alpha\beta}(p)$
are given in momentum space by 
\be\label{eqAI}
I_{\mu\nu , \alpha\beta} = \frac 1 2 (\eta_{\mu\alpha}\eta_{\nu\beta} 
-\eta_{\nu\alpha}\eta_{\mu\beta} ) 
\ee 
and 
\be\label{eqAL}
L_{\mu\nu ,\alpha\beta}(p)=\frac 1 2\left(
p_\mu p_\alpha\eta_{\nu\beta}+p_\nu p_\beta\eta_{\mu\alpha}-
p_\nu p_\alpha\eta_{\mu\beta}-p_\mu p_\beta\eta_{\nu\alpha}
\right).
\ee
It is also convenient to denote
the free propagators in 
Eqs. \eqref{eqA1a}, \eqref{eqA1b}, 
\eqref{eqA1c} and \eqref{eqA1d} 
respectively as  
${D^{(0)}_{FF}}_{\mu\nu , \alpha\beta}^{ab}(p)$, 
${D^{(0)}_{AA}}_{\mu  \nu}^{ab}(p)$, 
${D^{(0)}_{AF}}_{\mu , \alpha\beta}^{ab}(p)$ and 
${D^{(0)}_{FA}}_{\alpha\beta , \mu}^{ab}(p)$.

Note that the tensors \eqref{eqAI} and \eqref{eqAL}  satisfy 
\be\label{kIkL}
p^\rho I_{\mu\nu , \alpha\rho} = p^\rho \frac{1}{p^2} L_{\mu\nu , \alpha\rho}(p) =
\frac 1 2\left(\eta_{\mu\alpha} p_\nu - \eta_{\nu\alpha} p_\mu\right) 
\ee
which imply that the $F$-propagator in \eqref{eqA1} satisfies the transversality
condition
\be\label{transvFF1}
p^\mu {D^{(0)}_{FF}}_{\mu\nu , \alpha\beta}^{ab}(p) = 0.
\ee
Also, the identities
\be\label{idemp1}
I_{\mu\nu , \lambda\rho} \frac{1}{p^2} L^{\lambda\rho}_{\;\;\;\;\alpha\beta}(p) =
\frac{1}{p^2} L_{\mu\nu , \lambda\rho}(p)
\frac{1}{p^2} L^{\lambda\rho}_{\;\;\;\;\alpha\beta}(p) =
\frac{1}{p^2} L_{\mu\nu, \alpha\beta}(p) 
\ee
($L_{\mu\nu , \alpha\beta}(p)/p^2$ is idempotent)
imply that the $F$-propagator satisfies the relation
\be
L_{\mu\nu, \lambda\rho}(p) {D^{(0)}_{FF}}^{\lambda\rho\, ab}_{\;\;\;\; \alpha\beta}(p)=0. 
\ee

For completeness, let us also display the well 
known Feynman rules obtained from the second-order formalism 
Lagrangian given by Eq. \eqref{eq11}. The propagators for the
$A_\mu^a$ and the ghost fields, as well as the ghost vertex, are the
same as in Eqs.  \eqref{eqA1b}, \eqref{eqA1f} and \eqref{eqA1g}. But
now, instead of the single momentum independent vertex, given by
\eqref{eqA1e}, as well as the mixed propagator in Eq. \eqref{eqA1c}, 
we have the following cubic and quartic vertices
\begin{subequations}
\begin{eqnarray}
\begin{array}{c}\label{eqA4a} 
\input{three_gluon.pspdftex} 
\end{array} \;\;\;\;  &\;\;& 
{g}f^{abc}\left[\left(p_\nu-q_\nu\right)\eta_{\mu\alpha}+
\left(q_\alpha-k_\alpha\right)\eta_{\mu\nu}+
\left(k_\mu-p_\mu\right)\eta_{\alpha\nu}
\right], 
\\
\nonumber 
\\
\begin{array}{c}\label{eqA4b} 
\input{four_gluon.pspdftex} 
\end{array} \;\;\;\;  &\;\;& 
\begin{array}{l}
\displaystyle{ -ig^2\bigg[
f^{abe} f^{cde} (\eta_{\alpha\mu} \eta_{\beta\nu} - \eta_{\beta\mu} \eta_{\alpha\nu} )+}\\
\;\;\;\;\;\;\;\;\;\;\displaystyle{
f^{ace} f^{bde} (\eta_{\alpha\beta} \eta_{\mu\nu} - \eta_{\beta\mu}\eta_{\alpha\nu} )+}\\
\;\;\;\;\;\;\;\;\;\;\displaystyle{
f^{bce} f^{ade} (\eta_{\alpha\beta} \eta_{\mu\nu} - \eta_{\alpha\mu} \eta_{\beta\nu} ) 
\bigg]} , 
\end{array}
\end{eqnarray}
\end{subequations}
with all the momenta are flowing inwards.

The identities  \eqref{eq19} and  \eqref{eq13}
can be verified at the lowest
order, in the momentum space, using the free 
propagators introduced in the Feynman rules above.
The tree level momentum space version of Eq. \eqref{eq19} 
\be\label{treeFA1}
{D^{(0)}_{FA}}_{\lambda\sigma , \nu}^{ab}
=i p_\lambda {D^{(0)}_{AA}}_{\sigma \nu}^{ab} -
i p_\sigma {D^{(0)}_{AA}}_{\lambda\nu}^{ab} 
\ee
is verified using Eqs. \eqref{eqA1d} and \eqref{eqA1b}
(note that the momentum space 
expression of the bi-linears like $(\partial A(x) \cdots  A(x))$ is
$ip \tilde A(p) \cdots \tilde A(-p)$).

Similarly, the momentum space form of Eq. \eqref{eq13} can be written as 
\be
{D^{(0)}_{FF}}_{\mu\nu , \alpha\beta}^{ab}
= 2 i \delta^{ab} I_{\mu\nu, \alpha\beta} +
 p_\mu p_\alpha {D^{(0)}_{AA}}_{\nu \beta}^{ab} 
-p_\mu p_\beta  {D^{(0)}_{AA}}_{\nu \alpha}^{ab} 
-p_\nu p_\alpha {D^{(0)}_{AA}}_{\mu \beta}^{ab} 
+p_\nu p_\beta {D^{(0)}_{AA}}_{\mu \alpha}^{ab}, 
\ee
which can be readily verified using the Eqs. \eqref{eqA1a} and
\eqref{eqA1b}.


\section{One-loop results}

\subsection{Self-energies}

\begin{figure}[t]
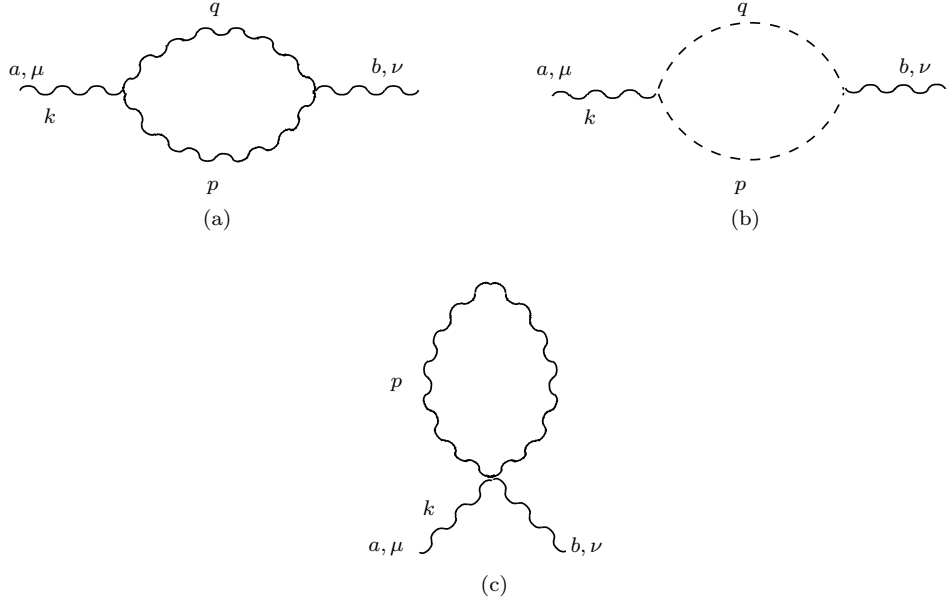

\begin{eqnarray}
\input{propAA_2nd_a.pspdftex} & \qquad \qquad  
\input{propAA_2nd_b.pspdftex} 
 \nonumber 
\end{eqnarray}
\begin{eqnarray}
   \input{propAA_2nd_c.pspdftex}
\nonumber 
\end{eqnarray}
\caption{One-loop contributions to the gauge field self-energy in the  
  second-order formalism 
($q=p+k$ and $k$ in an inward external momentum).  
There is a combinatorial $1/2$
  factor associated with diagrams (a) and (c) and a minus sign  
  associated with the ghost-loop diagram in (b).  
The relevant Feynman 
  rules are given in Eqs. \eqref{eqA1b}, \eqref{eqA1f}, \eqref{eqA1g}, 
\eqref{eqA4a} 
and \eqref{eqA4b}.}
\label{fig2}
\end{figure}

\subsubsection{The general method and the $A_\mu^a$ self-energy in the second-order formalism}
Let us first consider all the possible self-energy diagrams that can
be computed using the Feynman rules presented in Appendix A.  
As is well known, these basic 1PI diagrams are the basic building 
blocks that contribute to the identities like the ones 
given by Eqs. \eqref{eq19} and \eqref{eq13}. 

\begin{figure}[h]
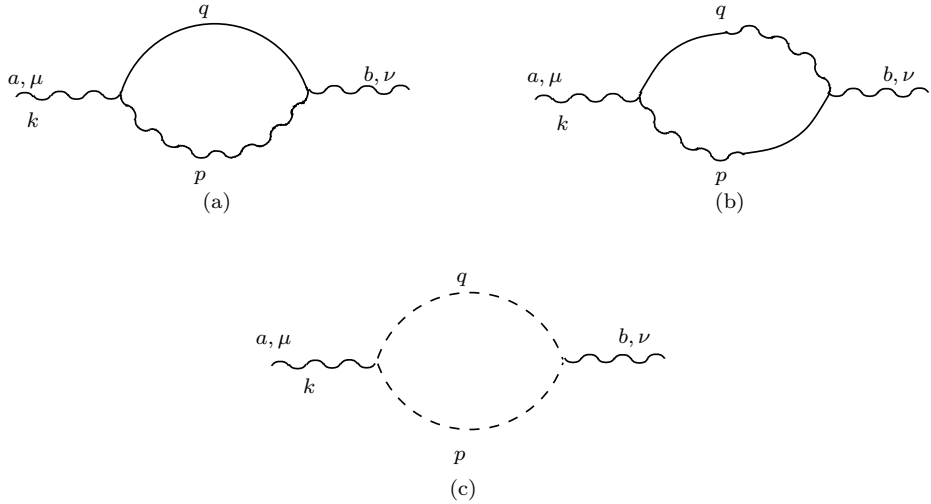

\begin{eqnarray}
\input{AAse_1st_a.pspdftex}  & \qquad \qquad 
\input{AAse_1st_b.pspdftex} 
 \nonumber 
\end{eqnarray}
\begin{eqnarray}
   \input{AAse_1st_c.pspdftex}
\nonumber 
\end{eqnarray}
\caption{One-loop contributions to the gauge field self-energy in the
  first-order formalism 
($q=p+k$ and $k$ in an inward external momentum). 
The relevant Feynman rules are given in Eqs. \eqref{eqA1}.}
\label{fig3}
\end{figure}

The diagrams which contribute to the well known result for $A$ field self-energy, in the second-order
formalism, are shown in Fig. \ref{fig2} (diagrams in Figs. (2a), (2c) and (4c) have a combinatorial factor $1/2$;
there is a minus sign for the ghost loop diagrams). Fig. \ref{fig3} show the contributions to the $A$ field
self-energy in the first-order formalism.
Our basic approach for the computation of all the self-energies will be based on tensor
decompositions. In the case of Figs. \ref{fig2} and \ref{fig3}, all
the diagrams will have, after the loop momentum integration,
the following co-variant tensor structure
\be\label{gente1}
{\Pi^{I}}^{ab}_{\mu\nu}(k) = C_{YM} \delta^{ab} \left(C^I_1 \eta_{\mu\nu} 
+ C^I_2 \frac{k_\mu k_\nu}{k^2}\right)
; \;\;\; I=\mbox{(2a), }\mbox{(2b), }\mbox{(2c)}
\ee
(we are using $f^{amn} f^{bmn} = C_{YM} \delta^{ab}$).

The coefficients $C^I_i$ can be obtained solving the following system of two algebraic equations
\be\label{sys1}
\left\{\begin{array}{lll}
         \eta^{\mu\nu}  {\Pi^{I}}_{\mu\nu}(k) 
           &=&   d C^I_1  + C^I_2   \\ & & \\
          k^\mu k^\nu   {\Pi^{I}}_{\mu\nu}(k)  
             &=&    k^2 C^I_1  + k^2 C^I_2  
\end{array}\right. ;\;\;\; I=\mbox{(2a), }\mbox{(2b) and }\mbox{(2c)}  ,
\ee
where we have introduced ${\Pi^{I}}_{\mu\nu}(k)$ (without the colour indices) such that
${\Pi^{I}}^{ab}_{\mu\nu}(k) = C_{YM} \delta^{ab} {\Pi^{I}}_{\mu\nu}(k)$.

Using the Feynman rules given in Appendix A, 
the scalar integrals on the left hand side of the Eqs. \eqref{sys1} will have the following form
\be\label{interals1}
\int \frac{d^d p}{(2 \pi)^d}  s^I(p,q,k),
\ee
where $q=p+k$; $p$ is the loop momentum, $k$ is the external momentum and $s^I(p,q,k)$ are
scalar functions of $p\cdot k $, $q\cdot k $, $p\cdot q $, $p^2$,
$q^2$ and $k^2$. Upon using the following  relations 
\begin{subequations}
\begin{eqnarray}
p\cdot k = (q^2 - p^2 - k^2)/2, \\
q\cdot k = (q^2 + k^2 - p^2)/2, \\
p\cdot q = (p^2 + q^2 - k^2)/2, 
\end{eqnarray}
\end{subequations}
the scalars  $s^I(p,q,k)$ can be reduced to combinations of powers of $p^2$ and $q^2$. As a result, the
integrals in Eq. \eqref{interals1}
can all be expressed in terms of combinations of the following simple integrals 
\begin{equation}
I^{l m} \equiv 
\int \frac{d^d p}{(2 \pi)^d} \frac{1}{(p^2)^l (q^2)^m} = i^{d+1} \frac{(k^2)^{d/2-l-m}}{(4\pi)^{d/2}}
\frac{\Gamma(l+m-d/2)}{\Gamma(l) \Gamma(m)} \frac{\Gamma(d/2-l) \Gamma(d/2-m)}{\Gamma(d-l-m)},
\end{equation}
where  powers $l> 1$ and $m> 1$ may only arise from the terms proportional to $1-\xi$ in the gluon propagator (see Eq. \eqref{eqA1b}). 
The only non-vanishing (i.e. non tadpole) integrals are 
\begin{subequations}\label{intregd}
\begin{eqnarray}
I^{11} & = & i^{d+1} \frac{(k^2)^{d/2-2}}{2^d\pi^{d/2}}
\frac{\Gamma \left(2-\frac{d}{2}\right) \Gamma \left(\frac{d}{2}-1\right)^2}{\Gamma (d-2)} \\
I^{12} & = & I^{21} = \frac{(3-d) } {k^2}  I^{11}\\ 
I^{22} & = & \frac{(3-d) (6-d) } {k^4}  I^{11}.
\end{eqnarray}
\end{subequations}
In $d=4-2\epsilon$ dimensions, $1/\epsilon$ 
ultra-violet pole part of the basic integral  $I^{11}$ is given by
\be\label{poleI11}
I^{11} = \frac{i}{16\pi^2\epsilon}.
\ee

Implementing the above-described procedure as a straightforward
computer algebra code, we readily obtain the following results
\begin{subequations}
\be\label{se2nda}
\left\{\begin{array}{lll}
C^{{(2a)}}_1 &=\displaystyle{
-\frac{k^{2} \left(d^2 (\xi -1) (\xi +7)+d (19-\xi  (5 \xi +26))+4 \xi  (\xi
   +5)-14\right)}{8 (d-1)}} g^2 I^{11}
& \\ && \\
C^{{(2a)}}_2 &=-C^{{(2a)}}_1
               -\displaystyle{\frac{k^2}{4}} g^2 I^{11}& 
\end{array}\right . ,
\ee
\be\label{se2ndb}
\left\{\begin{array}{lll}
C^{{(2b)}}_1 &=\displaystyle{\frac{k^2}{4(d-1)}} g^2 I^{11}
& \\ && \\
C^{{(2b)}}_2 &=\displaystyle{\frac{k^2}{4}} g^2 I^{11} - C^{(2b)}_1
& 
\end{array}\right . ,
\ee
\be\label{se2ndc}
C^{{(2c)}}_1  
=C^{{(2c)}}_2 =0 ,  
\ee
\end{subequations}
where we have used the formulas in Eqs. \eqref{intregd}.

Eq. \eqref{se2ndc}  follows from the tadpole nature of the
diagram in Fig. (2c) which vanishes when one uses dimensional regularization.
Notice that Eqs. \eqref{se2nda}, \eqref{se2ndb} and \eqref{se2ndc} imply that 
$C^{(2a)}_2 + C^{(2b)}_2 + C^{(2c)}_2 = -(C^{(2a)}_1 + C^{(2b)}_1 + C^{(2c)}_1)$,
so that the self-energy will be transverse. The sum
$C^{(2a)}_1 + C^{(2b)}_1 + C^{(2c)}_1$ gives the following result
\be
\frac{d^2 (1-\xi) (\xi +7)+d (\xi  (5 \xi +26)-19)-4 \xi  (\xi +5)+16}{8 (d-1)}
g^2 I^{11} k^2 .
\ee
Therefore, the final result for the $A_\mu^a$ field self-energy in the
second-order formalism is given by 
\begin{eqnarray}\label{se2nd}
{\Pi_{AA}^{{(2nd)}}}_{\mu\nu}^{ab}(k) &=&
                                     {\Pi^{{(2a)}}}_{\mu\nu}^{ab}(k)+{\Pi^{{(2b)}}}_{\mu\nu}^{ab}(k)+{\Pi^{{(2c)}}}_{\mu\nu}^{ab}(k) 
                                     \nonumber \\
&=& 
\frac{d^2 (1-\xi) (\xi +7)+d (\xi  (5 \xi +26)-19)-4 \xi  (\xi +5)+16}{8 (d-1)}
g^2 I^{11} 
C_{YM} \delta^{ab} \left(k^2\eta_{\mu\nu} - k_\mu k_\nu\right) 
\end{eqnarray}
which is in agreement with the well known result in $d$ dimensions
(see Eq. (A.12) of \cite{muta:book87} and the comment on the
missing factor of $i$ on page 81).
Using \eqref{poleI11}
we obtain the following  UV pole part
($g^2 I^{11} \approx {i g^2}/(16 \pi^2 \epsilon )$)
\be\label{Ase2ndUV}
{\Pi_{AA}^{{(2nd)}}}_{\mu\nu}^{ab}(k) = i \frac{C_{YM} g^2}{16\pi^2\epsilon} 
\frac{13-3\xi}{6} \delta^{ab}
\left(k^2\eta_{\mu\nu} - k_\mu k_\nu\right) + \cdots, 
\ee
which is in agreement with the well known result (see Eqs. (A.19) and
(A.21) of \cite{muta:book87}).

\subsubsection{$A_\mu^a$ self-energy in the first-order formalism}
Let us now consider the $A_\mu^a$ self-energy in the first-order
formalism. The one-loop diagrams are shown in Fig \ref{fig3}.
Using the Feynman rules given in the Appendix A and 
considering that we have the same co-variant structure as in
Eq. \eqref{gente1}, Eqs. \eqref{sys1} can be solved, with
$I=\mbox{(3a), }\mbox{(3b) and }\mbox{(3c)}$, yielding
following results
\begin{subequations}
\be\label{se1sta}
\left\{\begin{array}{lll}
C^{{(3a)}}_1 &=\displaystyle{\frac{k^2(d-2)}{4(d-1)}}\left[
d(1-\xi)+\xi-2
\right] g^2 I^{11}
& \\ && \\
C^{{(3a)}}_2 &= \displaystyle{\frac{k^2}{4}(d-2)} g^2 I^{11} -C_1^{(3a)}  
& 
\end{array}\right . ,
\ee
\be\label{se1stb}
\left\{\begin{array}{lll}
C^{{(3b)}}_1 &=\displaystyle{-\frac{k^2}{4}} g^2 I^{11}& \\ && \\
C^{{(3b)}}_2 &=\displaystyle{-\frac{k^2}{4}}(d-1) g^2 I^{11} - C^{(3b)}_1& 
\end{array}\right . ,
\ee
\be\label{se1stc}
\left\{\begin{array}{lll}
C^{{(3c)}}_1 &=& \displaystyle{\frac{k^2}{4(d-1)}} g^2 I^{11} \\ && \\
C^{{(3c)}}_2 &=& \displaystyle{\frac{k^2}{4}} g^2 I^{11} - C^{(3c)}_1 
\end{array}\right . ,
\ee
\end{subequations}
where we have used Eqs. \eqref{intregd}.

As we can see the transversality condition is also satisfied 
in the first-order formalism. Indeed, Eqs. \eqref{se1sta}, \eqref{se1stb} and
\eqref{se1stc}, imply that 
$C^{(3a)}_2 + C^{(3b)}_2 + C^{(3c)}_2 = -(C^{(3a)}_1 + C^{(3b)}_1 +C^{(3c)}_1)$.
From the sum $C^{(3a)}_1 + C^{(3b)}_1 +C^{(3c)}_1$ we obtain the following result
for the $A_\mu^a$ self-energy in the first-order formalism
\begin{eqnarray}\label{se1st}
{\Pi_{AA}^{{(1st)}}}_{\mu\nu}^{ab}(k) &=&
                                     {\Pi_{AA}^{{(3a)}}}_{\mu\nu}^{ab}(k)+{\Pi_{AA}^{{(3b)}}}_{\mu\nu}^{ab}(k)+{\Pi_{AA}^{{(3c)}}}_{\mu\nu}^{ab}(k) 
                                     \nonumber \\
&=& 
\displaystyle{\frac{(d-2)\left[ d(1-\xi)+\xi-3\right]}{4(d-1)}} g^2 I^{11} 
C_{YM} \delta^{ab} \left(k^2\eta_{\mu\nu} - k_\mu k_\nu\right) 
\end{eqnarray}
which is different from the result in the second-order formalism,
given by Eq. \eqref{se2nd}. 
Using \eqref{poleI11}
we obtain the following result for the UV pole
\be\label{Ase1stUV}
{\Pi_{AA}^{{(1st)}}}_{\mu\nu}^{ab}(k) = 
i \frac{C_{YM} g^2}{16\pi^2\epsilon} 
\frac{1-3\xi}{6} \delta^{ab}
\left(k^2\eta_{\mu\nu} - k_\mu k_\nu\right) + \cdots.
\ee
Of course there is no contradiction
with the general conclusions of section II, which asserts that
the two formalisms should have 
the same Green's functions containing only 
external $A_\mu^a$ fields; the 1PI functions are not necessarily the
same in both formalisms. On the other hand, 
as we will see bellow from Eq. \eqref{propAA}, 
the propagator for the $A_\mu^a$ field is the same in both formalisms. 

\subsubsection{$F_{\mu\nu}^a$ self-energy}\label{secB1c}
\begin{figure}[b]
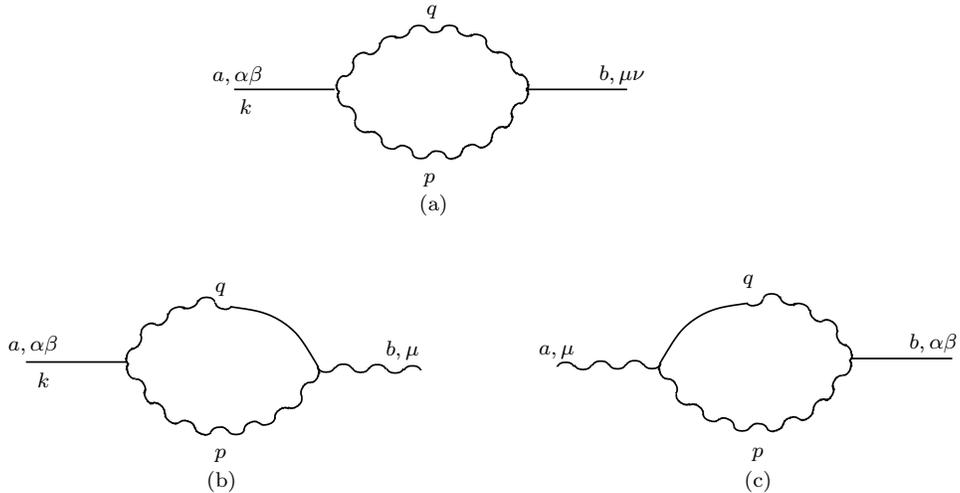

\begin{eqnarray}
\input{FFse.pspdftex}  & \qquad \qquad  
\nonumber 
\end{eqnarray}
\begin{eqnarray}
\input{FAse_1st.pspdftex}  & \qquad \qquad 
\input{AFse_1st.pspdftex} 
 \nonumber 
\end{eqnarray}
\caption{One-loop contributions to the self-energies 
with external $F$ fields in the first-order formalism 
($q=p+k$ and $k$ in an inward external momentum). 
The relevant Feynman 
  rules are given in Eqs. \eqref{eqA1}.}
\label{fig4}
\end{figure}

A complete tensor basis for
the diagram in Fig. (\ref{fig4}a) 
can be formed using the two tensors introduced
in Eqs. \eqref{eqAI} and \eqref{eqAL}, since these are the most
general tensors with four indices and having the required symmetry.
Similarly to Eq. \eqref{gente1} we can write
\be\label{seFF1}
{\Pi_{FF}^{(4a)}}_{\alpha\beta ,\mu\nu }^{ab}(k) = C_{YM}\delta^{ab}
\left[
C_1^{(4a)} I_{\alpha\beta ,\mu\nu} 
+ C_2^{(4a)} \frac{1}{k^2} L_{\alpha\beta , \mu\nu}(k)\right]. 
\ee 
Proceeding similarly as in the case of Eq. \eqref{sys1}, we contract
the Eq. \eqref{seFF1} with $I_{\mu\nu,\alpha\beta}$ and 
$L_{\mu\nu,\alpha\beta}(k)$ and solve the system of equations for
$C_1^{(4c)}$ and $C_2^{(4c)}$ (as in the previous calculations,
this is a very straightforward and well
defined computer algebra procedure), yielding the following
results
\be\label{seFFa}
\left\{\begin{array}{lll}
C^{(4a)}_1 &=& \displaystyle{\frac{\xi+1}{4}} g^2 I^{11} \\ && \\
C^{(4a)}_2 &=& -\frac{1}{16}(d-4)(\xi^2-1) g^2 I^{11}
\end{array}\right . . 
\ee
Using \eqref{poleI11}
(notice that 
$C_2^{(4a)}$ does not have a pole $1/\epsilon$), 
we obtain the following result for the UV pole
\be\label{seFFUV}
{\Pi_{FF}^{(4a)}}_{\alpha\beta ,\mu\nu}^{ab}(k) = 
i \frac{C_{YM} g^2}{16\pi^2\epsilon} 
\displaystyle{\frac{\xi+1}{4}}  \delta^{ab} 
I_{\alpha\beta , \mu\nu}  
+ \cdots .
\ee

\subsubsection{$FA$ and $AF$ self-energies}\label{secFA}
Finally let us consider the mixed self-energies in Figs. (4b) and (4c).
In this case, there is just one tensor with three indices and the
required symmetry. For instance, in the case of the diagram in
Fig. (\ref{fig4}b), the result can be expressed as follows
\be\label{seFAa}
{\Pi_{FA}^{(4b)}}_{\alpha\beta ,\mu}^{ab}(k) = 
C^{(4b)} C_{YM}
\delta^{ab} 
\left(k_\alpha\eta_{\beta\mu}-k_\beta\eta_{\alpha\mu}\right).
\ee
Contracting both sides with the Lorentz structure on the right hand
side and performing the integrals with the help of \eqref{intregd}, we
obtain
\be
C^{(4b)} = \frac{i}{8}\left[d(1-\xi)+3 \xi-1\right] g^2 I^{11}.
\ee
Using \eqref{poleI11}
we obtain the following result for the UV pole
\be\label{seFAaUV}
{\Pi_{FA}^{(4b)}}_{\alpha\beta ,\mu}^{ab}(k) = 
\frac{C_{YM} g^2}{16\pi^2\epsilon} \delta^{ab}  \frac{\xi-3}{8} 
\left(k_\alpha\eta_{\beta\mu}-k_\beta\eta_{\alpha\mu}\right).
\ee
Proceeding similarly, we obtain
\be\label{seAFa}
{\Pi_{AF}^{(4c)}}_{\mu,\alpha\beta}^{ab}(k) = -{\Pi_{FA}^{(4b)}}_{\alpha\beta,\mu}^{ab}(k) .
\ee

\subsection{Propagators}
\subsubsection{The $FF$ propagator}
Using the results for the self-energies, the propagator for the
$F_{\mu\nu}^a$ field can be expressed as follows
\begin{eqnarray}\label{propFF1}
{D^{(1)}_{FF}}_{\mu \nu , \alpha\beta}^{ab} &=& 
{D^{(0)}_{FA}}_{\mu\nu,  \rho}^{ac} 
{\Pi_{AA}^{{(1st)}}}^{cd\,  \rho\lambda} 
{D^{(0)}_{AF}}_{\lambda, \alpha\beta}^{db} 
+
{D^{(0)}_{FF}}_{\mu\nu,  \rho\sigma}^{ac} 
{\Pi_{FF}^{(4a)}}^{cd\,  \rho\sigma , \lambda\delta} 
{D^{(0)}_{FF}}_{\lambda\delta , \alpha\beta}^{db} 
\nonumber \\
&+&
{D^{(0)}_{FF}}_{\mu\nu,  \rho\sigma}^{ac} 
{\Pi_{FA}^{(4b)}}^{cd\,  \rho\sigma , \gamma} 
{D^{(0)}_{AF}}_{\gamma, \alpha\beta}^{db} 
+
{D^{(0)}_{FA}}_{\mu\nu  ,  \rho}^{ac} 
{\Pi_{AF}^{(4c)}}^{cd\,  \rho, \gamma\delta} 
{D^{(0)}_{FF}}_{\gamma\delta, \alpha\beta}^{db} .
\end{eqnarray}
From the properties of the tensors $L_{\mu\nu, \alpha\beta}$, 
$(k_\mu \eta_{\nu\alpha} - k_\nu \eta_{\mu\alpha})$
as well as the identity $I_{\mu\nu, \alpha\beta}$ 
one can show that the last two terms 
in Eq. \eqref{propFF1} vanish
(see Eqs. \eqref{kIkL}, \eqref{transvFF1} and \eqref{idemp1}).
Using the results for
${\Pi_{AA}^{{(1st)}}}^{cd\,  \alpha\beta}$ and 
${\Pi_{FF}^{(4a)}}^{cd\,  \alpha\beta , \gamma\delta}$
given respectively in Eqs. \eqref{se1st} and \eqref{seFF1}
a straightforward calculation yields
\be \label{propFF2}
{D^{(1)}_{FF}}_{\mu \nu , \alpha\beta}^{ab}  =
g^2 I^{11}C_{YM} \delta^{ab} \left\{
-(\xi+1) I_{\mu\nu, \alpha\beta}+
\displaystyle{\frac{d\left[d(\xi-1)-\xi+7\right]-8}{2(d-1) k^2}}L_{\mu\nu,\alpha\beta}(k) 
\right\}.
\ee
Using \eqref{poleI11}
we obtain the following  UV pole part
\be \label{propFFUV}
{D^{(1)}_{FF}}_{\mu \nu , \alpha\beta}^{ab}  =
i \frac{C_{YM} g^2}{16\pi^2\epsilon} \delta^{ab}
\left\{
-(\xi+1) I_{\mu\nu, \alpha\beta}+
\left(\frac{2}{3} + 2 \xi\right) 
\frac{1}{k^2}L_{\mu\nu,\alpha\beta}(k) 
\right\} + \cdots .
\ee

\subsubsection{The $AA$ propagator}
The one-loop correction to the $A_\mu^a$ propagator can be calculated from
\begin{eqnarray}\label{propAA1}
{D^{(1)}_{AA}}_{\mu \nu}^{ab} &=& 
{D^{(0)}_{AA}}_{\mu  \alpha}^{ac} 
{\Pi_{AA}^{{(1st)}}}^{cd\,  \alpha\beta} 
{D^{(0)}_{AA}}_{\beta  \nu}^{db} 
+
{D^{(0)}_{AF}}_{\mu,  \alpha\beta}^{ac} 
{\Pi_{FF}^{(4a)}}^{cd\,  \alpha\beta , \gamma\delta} 
{D^{(0)}_{FA}}_{\gamma\delta, \nu}^{db} 
\nonumber \\
&+&
{D^{(0)}_{AF}}_{\mu,  \alpha\beta}^{ac} 
{\Pi_{FA}^{(4b)}}^{cd\,  \alpha\beta , \gamma} 
{D^{(0)}_{AA}}_{\gamma\nu}^{db} 
+
{D^{(0)}_{AA}}_{\mu  \alpha}^{ac} 
{\Pi_{AF}^{(4c)}}^{cd\,  \alpha , \gamma\delta} 
{D^{(0)}_{FA}}_{\gamma\delta, \nu}^{db} . 
\end{eqnarray}
Using the results for 
${\Pi_{AA}^{{(1st)}}}^{cd\,  \alpha\beta}$, 
${\Pi_{FF}^{(4a)}}^{cd\,  \alpha\beta , \gamma\delta}$, 
${\Pi_{FA}^{(4b)}}^{cd\,  \alpha\beta , \gamma}$ and 
${\Pi_{AF}^{(4c)}}^{cd\,  \alpha , \gamma\delta}$ 
given respectively in Eqs. \eqref{se1st}, \eqref{seFF1}, \eqref{seFAa}
and \eqref{seAFa}, a straightforward calculation yields
\be \label{propAA}
{D^{(1)}_{AA}}_{\mu \nu}^{ab} = -
 \frac{d^2 (1-\xi) (\xi +7)+d (\xi  (5 \xi +26)-19)-4 \xi  (\xi +5)+16}{8 (d-1)}
g^2 I^{11} 
C_{YM} \delta^{ab} 
\left(\frac{\eta_{\mu\nu}}{k^2} - \frac{k_\mu k_\nu}{k^4}\right) . 
\ee
It is immediately clear that Eq. \eqref{propAA}  is the same as the
propagator for the $A_\mu^a$ field in the second-order formalism which
can be obtained by simply computing
${D^{(0)}_{AA}}_{\mu  \alpha}^{ac} 
{\Pi_{AA}^{{(2nd)}}}^{cd\,  \alpha\beta} 
{D^{(0)}_{AA}}_{\beta  \mu}^{db} $, where
${\Pi_{AA}^{{(2nd)}}}^{cd\,  \alpha\beta}$ is given by Eq. \eqref{se2nd}
(using the transversality, it is easy to see that this will just produce
a factor $(-1/k^4)$ times the self-energy in Eq. \eqref{se2nd}).
This is an explicit special example of the
general result, pointed out in section II, according to which
the two formalisms give the same Green's functions containing only 
external $A_\mu^a$ fields, for any choice of the gauge parameter and
dimension $d$. 
Using \eqref{poleI11}
in Eq. \eqref{Ase2ndUV}, we obtain the following UV pole part
\be\label{propAAUV1}
{D^{(1)}_{AA}}_{\mu \nu}^{ab} 
=i \frac{C_{YM} g^2}{16\pi^2\epsilon}    
\frac{3\xi-13}{6} \delta^{ab}
\left(\frac{\eta_{\mu\nu}}{k^2} - \frac{k_\mu k_\nu}{k^4}\right) + \cdots. 
\ee

Using the result for the $A_\mu^a$ propagator in Eq. \eqref{propAA},  we can now 
compute the quantity 
\be\label{linff}
\langle 0 |T (\partial_\mu A_\nu^a - \partial_\nu A_\mu^a)(x) 
(\partial_\alpha A_\beta^b - \partial_\beta A_\alpha^b)(y)| 0\rangle , 
\ee
which is part of the contribution to the right hand side of 
Eq. \eqref{eq13}.
The corresponding expression 
in momentum space  ($\partial_\mu\rightarrow i k_\mu$ 
for the first  momentum and 
$\partial_\mu\rightarrow -i k_\mu$ for the second momentum) 
is given by
\be
 k_\mu k_\alpha {D^{(1)}_{AA}}_{\nu \beta}^{ab} 
-k_\mu k_\beta  {D^{(1)}_{AA}}_{\nu \alpha}^{ab} 
-k_\nu k_\alpha {D^{(1)}_{AA}}_{\mu \beta}^{ab} 
+k_\nu k_\beta {D^{(1)}_{AA}}_{\mu \alpha}^{ab}, 
\ee
where we are using that the $A_\mu^a$ 
propagator is the same in both formalisms.
Using Eq. \eqref{propAA} we obtain 
\be\label{dada1} 
-\frac{d^2 (1-\xi) (\xi +7)+d (\xi  (5 \xi +26)-19)-4 \xi  (\xi +5)+16}{8 (d-1)}
g^2 I^{11} 
C_{YM} \delta^{ab} \frac{1}{k^2}
\left(
k_\mu k_\alpha\eta_{\nu\beta}+k_\nu k_\beta\eta_{\mu\alpha}-
k_\nu k_\alpha\eta_{\mu\beta}-k_\mu k_\beta\eta_{\nu\alpha}
\right) 
\ee
(notice that terms like $k_\alpha k_\beta k_\mu k_\nu$ vanish due to
the anti-symmetry). Using Eq. \eqref{eqAL}, this can be written as
\be\label{dAdAg} 
 \frac{d^2 (\xi-1) (\xi +7)-d (\xi  (5 \xi +26)+19)+4 \xi  (\xi +5)-16}{4 (d-1)}
g^2 I^{11} 
C_{YM} \delta^{ab} \frac{1}{k^2}  L_{\mu\nu,\alpha\beta}(k),
\ee
which has the following UV pole
\be\label{dAdAUV}
i \frac{C_{YM} g^2}{16\pi^2\epsilon} 
\delta^{ab} 
\frac{3\xi-13}{3} 
\frac{1}{k^2}  L_{\mu\nu,
  \alpha\beta}(k). \ee

The $A_\mu^a$ propagator in Eq. \eqref{propAA} can also be used to 
compute the quantity
\be\label{linf} 
\langle 0 |T (\partial_\mu A_\nu^a - \partial_\nu A_\mu^a)(x) 
A^b_\alpha(y)| 0\rangle , 
\ee
that appears on the right hand side to the identity in Eq. \eqref{eq19}.
In momentum space, 
($\partial_\mu\rightarrow i k_\mu$)
this becomes
\be
i k_\mu {D^{(1)}_{AA}}_{\nu\alpha}^{ab} -
i k_\nu  {D^{(1)}_{AA}}_{\mu\alpha}^{ab}. 
\ee
Using Eq. \eqref{propAA}, we obtain
\be\label{adAtot} 
i \frac{d^2 (\xi-1) (\xi +7)
-d (\xi  (5 \xi +26)-19)+4 \xi  (\xi +5)-16}{8 (d-1)}
g^2 I^{11} 
C_{YM} \delta^{ab} 
\frac{1}{k^2}
\left(k_\mu \eta_{\nu\alpha} - k_\nu \eta_{\mu\alpha} \right) ,
\ee
which has the following UV pole part
\be\label{dAAUV}
\frac{C_{YM} g^2}{16\pi^2\epsilon}
\frac{13-3\xi}{6}
\frac{1}{k^2}
\left(k_\mu \eta_{\nu\alpha} - k_\nu \eta_{\mu\alpha} \right) 
+\cdots .
\ee

\subsubsection{The $FA$ and $AF$ propagators}
The one-loop contribution to the $FA$ propagator is given by
\begin{eqnarray}\label{propFA1}
{D^{(1)}_{FA}}_{\mu \nu , \alpha}^{ab} &=& 
{D^{(0)}_{FF}}_{\mu\nu,  \rho\sigma}^{ac} 
{\Pi_{FF}^{(4a)}}^{cd\,  \rho\sigma , \lambda\delta} 
{D^{(0)}_{FA}}_{\lambda\delta , \alpha}^{db} 
+
{D^{(0)}_{FF}}_{\mu\nu,  \rho\sigma}^{ac} 
{\Pi_{FA}^{(4b)}}^{cd\,  \rho\sigma , \gamma} 
{D^{(0)}_{AA}}_{\gamma, \alpha}^{db} 
\nonumber \\
&+&
{D^{(0)}_{FA}}_{\mu\nu,  \rho}^{ac} 
{\Pi_{AA}^{{(1st)}}}^{cd\,  \rho\lambda}  
{D^{(0)}_{AA}}_{\lambda, \alpha}^{db} 
+
{D^{(0)}_{FA}}_{\mu\nu  ,  \rho}^{ac} 
{\Pi_{AF}^{(4c)}}^{cd\,  \rho, \gamma\delta} 
{D^{(0)}_{FA}}_{\gamma\delta, \alpha}^{db} .
\end{eqnarray}
Using the results form the self-energies and the free propagators, the
first two terms in Eq. \eqref{propFA1} vanish
(see Eqs. \eqref{kIkL}, \eqref{transvFF1} and \eqref{idemp1})
and the sum of the last two terms yields
\be\label{propFA2}
{D^{(1)}_{FA}}_{\mu \nu , \alpha}^{ab} =i 
\displaystyle{\frac{\left[d(2d-7)(\xi-1)+5\xi-7\right]}{4(d-1)}}
g^2 I^{11} C_{YM} \delta^{ab}\frac{1}{k^2}
\left(k_\mu \eta_{\nu\alpha}-k_\nu \eta_{\mu\alpha}\right) .
\ee
Using \eqref{poleI11}
we obtain the following UV divergent result
\be\label{propFAUV}
{D^{(1)}_{FA}}_{\mu \nu , \alpha}^{ab} =
\frac{C_{YM} g^2}{16\pi^2\epsilon}
\delta^{ab}\frac{1}{k^2}
\frac{11-9\xi}{12}
\left(k_\mu \eta_{\nu\alpha}-k_\nu \eta_{\mu\alpha}\right) 
+\cdots .
\ee

Proceeding similarly, we obtain the following result for the $AF$ propagator 
\be\label{propAF2}
{D^{(1)}_{AF}}_{\alpha, \mu \nu }^{ab} =
-{D^{(1)}_{FA}}_{\mu \nu , \alpha}^{ab} 
\ee

\subsection{``Pinched'' diagrams}
Let us first consider the diagram in Fig (\ref{fig1}b). This arises
from the non-linear part 
$\langle 0|T f^{acd} A_\mu^c(x) A_\nu^d(x)  A_\alpha^b(y)  |0\rangle$
of $\langle 0|T f^a_{\mu\nu}(x)A^b_\alpha(y) |0\rangle$
%
(the linear part has been taken into 
account in Eq. \eqref{linf}, which has 
the momentum space result given 
in Eq. \eqref{adAtot}).
Since this contribution involves the product of two fields at the same
space-time point $x$, 
in the momentum space they become loop like diagrams, 
containing one cubic vertex, given by \eqref{eqA4a}, 
and two propagators given by \eqref{eqA1b}.
As we can see in Fig (\ref{fig6}),
the momentum structure is similar to the one shown
in the graph of Fig (\ref{fig4}b) but with no vertex 
on the left and replacing the $AF$ internal line by a $AA$ line.
Also, there is a free propagator on the right side.
Therefore, the corresponding expression 
in momentum space have the 
same tensor structure as in Eq. \eqref{eqA1d}.
Proceeding as in section (\ref{secFA}),  we find 
\be\label{eqAAAfA}
\frac{i}{8}
\left[
d(1-\xi)(\xi+3) + 2\xi(2\xi+5)-2\right]
g^2 I^{11}
C_{YM} \delta^{ab} 
\frac{1}{k^2}
\left(k_\mu \eta_{\nu\alpha}-k_\nu \eta_{\mu\alpha}\right) 
\ee
which has the following UV pole part
\be\label{eqAAAfAUV}
-\frac{C_{YM} g^2}{16\pi^2\epsilon} \delta^{ab}  \frac{\xi+5}{4} 
\frac{1}{k^2}
\left(k_\mu \eta_{\nu\alpha}-k_\nu \eta_{\mu\alpha}\right) 
+ \dots .
\ee
Adding the results in Eqs. \eqref{eqAAAfA} 
and \eqref{adAtot}, we obtain \eqref{propFA2}, 
which confirms the identity \eqref{eq19} to one loop
order.

\begin{figure}[t]
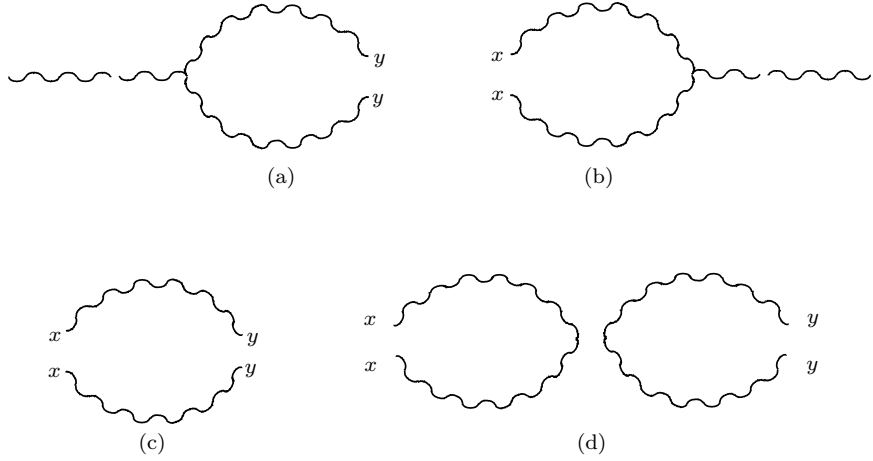

\begin{eqnarray}
\input{pinch_dAA_AAAAa.pspdftex}  & \qquad \qquad 
\input{pinch_dAA_AAAAb.pspdftex} 
 \nonumber 
\end{eqnarray}
\begin{eqnarray}
   \input{pinch_AAAAc.pspdftex} & \qquad \qquad 
\input{tadAAAA.pspdftex} 
\nonumber 
\end{eqnarray}
\caption{``Pinched'' contributions from 
    $\langle 0|T f^{acd} A_\mu^c(x) A_\nu^d(x) 
(\partial_\alpha A_\beta^b(y)  - 
\partial_\beta A_\alpha^b(y)) 
|0\rangle$ (a and b)
and
$\langle 0|T f^{acd} A_\mu^c(x) A_\nu^d(x)
f^{beg} A_\alpha^{e}(y) A_\beta^{g}(y) |0\rangle $ (c).
There is a second diagram identical to (c) which can be obtained
by interchanging the $x$ points.
The graphs in (d) vanish upon using dimensional regularization in
momentum space.}
\label{fig5}
\end{figure}

\begin{figure}[t]
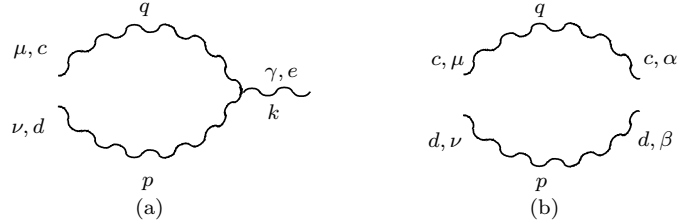

\begin{eqnarray}
\input{basicAAA.pspdftex}  & \qquad \qquad 
\input{basicAAAA.pspdftex}  
\nonumber
\end{eqnarray}
\caption{
The two basic momentum space graphs associated with Figs. 
(\ref{fig1}b) and  (\ref{fig5}). 
The momentum space expressions associated with
(\ref{fig1}b) and  (\ref{fig5}b) are 
obtained upon contracting (a) with
${g f^{acd} D^{(0)}_{AA}}_{\gamma\alpha}^{eb}(-k)$
and 
$g f^{acd}
\left[(-ik_\alpha){D^{(0)}_{AA}}_{\gamma\beta}^{eb}(-k) 
-\alpha\leftrightarrow \beta\right]$ respectively.
The momentum space expression associated with (\ref{fig5}c) 
is obtained  upon contracting (b) with $g^2 f^{acd} f^{bcd}$. 
}
\label{fig6}
\end{figure}

Fig. (\ref{fig5}) shows the contributions
from     $\langle 0|T f^{acd} A_\mu^c(x) A_\nu^d(x) 
(\partial_\alpha A_\beta^b(y) - \partial_\beta A_\alpha^b(y)) 
|0\rangle$ which arises from the 
non-linear parts of 
$\langle 0|T f_{\mu\nu}^a(x) f_{\alpha\beta}^b(y) |0\rangle$
(the linear part was considered 
in Eq. \eqref{linff}, which has 
the momentum space result given in Eq. \eqref{dAdAg}). 
Similarly to the previous calculation of the graphs in 
Fig. (\ref{fig1}b), the products of fields at the same space-time 
point $x$ give rise to loop integrals. 
As shown in Fig. (\ref{fig6}a), the momentum space expression
is a one-loop diagram similar to the contribution
from (\ref{fig1}b), but in this case the basic graph
is contracted with  
$g f^{acd}
\left[(-ik_\alpha){D^{(0)}_{AA}}_{\gamma\beta}^{eb}(k) 
-\alpha\leftrightarrow \beta\right]$. 
Since the corresponding expressions in momentum space have the 
same tensor structure as the $FF$-propagator, we can 
proceed as in  Sec. \ref{secB1c}. Using the same tensor
basis formed with the tensors 
in Eqs. \eqref{eqAI} and \eqref{eqAL}
we obtain for the sum of the contributions from 
Figs.  (\ref{fig1}a) and (\ref{fig1}b) the
following result
\be\label{fAAg}
g^2 I^{11} C_{YM} \delta^{ab}
\left[\frac 1 2 d(1-\xi)(\xi+3)+\xi(2\xi+5)-1\right]
\frac{1}{k^2} L_{\mu\nu , \alpha\beta}(k), 
\ee 
which has the following UV pole
\be\label{fAAUV}
i \frac{g^2 C_{YM}}{16\pi^2\epsilon} \delta^{ab}
(\xi+5) L_{\mu\nu , \alpha\beta}(k) + \cdots. 
\ee

Finally, we have the contribution from 
$\langle 0|T f^{acd}A^c_\mu(x)A^d_\nu(x) 
f^{beg}A^e_\alpha(y)A^g_\beta(y)|0\rangle$ 
in $\langle 0|T f^a_{\mu\nu}(x)f^b_{\alpha\beta}(y) |0\rangle$ 
which is  shown in Fig. (\ref{fig5}c)
(there is also an identical contribution obtained by interchanging 
the two $x$ points). 
Similarly to the previous cases, there is a momentum
space expression with a single loop associated with this contribution.
As before, the loop is associated with the pinch of the propagators at 
the same space-time point.  However, in this case we have
pinches at both sides so that there is no interaction vertex.
The corresponding loop diagram is shown in Fig. (\ref{fig6}b).
In terms of the tensors
$I_{\mu\nu ,\alpha\beta}$ and $L_{\mu\nu ,\alpha\beta}(k)$, we
obtain the following result for the loop integral
\be\label{AAAAg}
g^2 I^{11} C_{YM} \delta^{ab}\left[
-(\xi+1) I_{\mu\nu , \alpha\beta}
+\frac 1 4 (d-4)(\xi^2-1) \frac{1}{k^2} L_{\mu\nu , \alpha\beta}(k)
\right]
\ee
which has the following UV pole  
\be\label{AAAAUV}
-i \frac{g^2 C_{YM}}{16\pi^2} \delta^{ab}
(\xi+1) I_{\mu\nu , \alpha\beta}.
\ee

Adding the equations \eqref{dAdAg}, \eqref{fAAg}  
and \eqref{AAAAg}, a straightforward algebra 
shows that the result is the same as \eqref{propFF2}, so that 
the identity \eqref{eq13} is verified to one-loop order (of course
this remains true for the UV pole part).





\newpage


\begin{thebibliography}{10}

\bibitem{Okubo:1979gt}
S. Okubo and Y. Tosa, Phys. Rev. {\bf D20},  462  (1979), [Erratum: Phys.
  Rev.D23,1468(1981)].

\bibitem{McKeon:1994ds}
D.~G.~C. McKeon, Can. J. Phys. {\bf 72},  601  (1994). 

\bibitem{Martellini:1997mu}
M. Martellini and M. Zeni, Phys. Lett. {\bf B401},  62  (1997).

\bibitem{costello:2011b}
K. Costello, {\em Renormalisation and Effective Field Theory, Mathematical
  Surveys and Monographs} (American Mathematical Society, Providence, Rhode
  Island, 2011).

\bibitem{Brandt:2015nxa}
F.~T. Brandt and D.~G.~C. McKeon, Phys. Rev. {\bf D91},  105006  (2015).

\bibitem{Brandt:2016eaj}
F.~T. Brandt and D.~G.~C. McKeon, Phys. Rev. {\bf D93},  105037  (2016).

\bibitem{Frenkel:2017xvm}
J. Frenkel and J.~C. Taylor, Annals Phys. {\bf 387},  1  (2017).

\bibitem{Frenkel:2018xup}
J. Frenkel and J.~C. Taylor, Annals Phys. {\bf 389},  234  (2018).

\bibitem{Brandt:2018wxe}
F.~T. Brandt, J. Frenkel, and D.~G.~C. McKeon, Annals Phys. {\bf 409},  167932
  (2019).

\bibitem{Buchbinder:2018jqs}
I.~L. Buchbinder and P.~M. Lavrov, Eur. Phys. J. {\bf C78},  524  (2018).

\bibitem{Lavrov:2020exa}
P.~M. Lavrov, arXiv:2002.05997  (2020).

\bibitem{Barvinsky:2017zlx}
A.~O. Barvinsky, D. Blas, M. Herrero-Valea, S.~M. Sibiryakov, and C.~F.
  Steinwachs, JHEP {\bf 07},  035  (2018).

\bibitem{Wilson:1972ee}
K.~G. Wilson and W. Zimmermann, Commun. Math. Phys. {\bf 24},  87  (1972).

\bibitem{muta:book87}
T. Muta, {\em Foundations of Quantum Chromodynamics} (World Scientific,
  Singapore, 1987).

\bibitem{weinberg:book1995}
S. Weinberg, {\em Quantum Theory of Fields II} (Benjamin Cummings, Cambridge,
  1995).

\end{thebibliography}

\end{document}